\pdfoutput=1
\documentclass[aps,prl,preprint,groupedaddress,showpacs]{revtex4-1}
\usepackage{bm}
\usepackage{graphicx}
\usepackage{mathrsfs}
\usepackage{amssymb}
\usepackage{amsmath}
\usepackage{booktabs}
\usepackage{diagbox}
\usepackage{cases}

\begin{document}
\title{Property-aimed embedding: a machine learning frame work for material discovery}

\author{Lei Gu}
\affiliation{Department of Physics and Astronomy, University of California, Irvine, California 92697, USA}
\author{Ruqian Wu}
\email[]{wur@uci.edu}
\affiliation{Department of Physics and Astronomy, University of California, Irvine, California 92697, USA}
\begin{abstract}
Proposing new materials by atom substitution based on periodic table similarity is a conventional strategy of searching for materials with desired property. We introduce a machine learning frame work that promotes this paradigm to be property-specific and quantitative. It is of peculiar usefulness in situations where abundance data is accessible for learning general knowledge but samples for the problem of interest are relatively scarce. We showcase its usage and viability in the problem of separating high entropy alloys with different structural phases, for which a very simple data-driven criterion achieves differentiating ability comparable with widely used empirical criteria. Its flexibility and generability make it a promising tool in other material discovery tasks and far beyond.
\end{abstract}
\maketitle

\section{Introduction}
Owing to growing databases and algorithm improvements, machine learning (ML) techniques are achieving increasing capacity and popularity in material discovery~\cite{Butler2018,Sanchez2018,Zhuo2018,Ward2016,Gubernatis2018,Scherbela2018,Raccuglia2016,Ryan2018,Zhou2017,Jaeger2018,Schutt2017,Xu2019,Askera2019,Schutt2014,Xie2018,Xue2016,Li2017}. The core of these ML tasks is to construct a mapping between fundamental parameters and targeted properties, so that we can predict unexplored materials using data at hand and/or easy to acquire. Because the mapping in itself does not give us guidance for where to look for promising matters, we need clues from elsewhere. Atoms substitution based on their periodic table similarity is a conventional material discovery paradigm. Here, we present a framework that inherits and promotes the gist of this paradigm. Due to complex interactions among different ingredients, periodic table can not always be an effective guidance for the selection of substituent that provides the targeted property. With the proposed property-aimed embedding (PAE) approach, we can make this substitution paradigm much more property-specified and quantitative.

\begin{figure}
\includegraphics[width=0.45\textwidth]{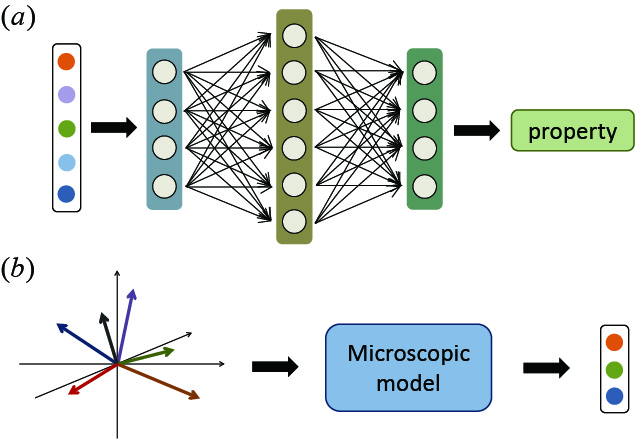}
\caption{(a) In PPM, a vector containing property informations of a matter is mapped to the targeted property. (b) In PAE, ingredient entities are embedded as vectors and mapped to relevant properties through a microscopic model/description.}
\label{rationale}
\end{figure}

It is convenient to have an idea of the PAE through comparison between its rationale and that of the property-to-property mapping (PPM) approach. In PPM, vectors encoding the property values are fed into a machine learning model (such as a neural network in Fig.~\ref{rationale}(a)), and a mapping is learned by optimizing the model variables. A classification problem is address similarly by changing the targeted property values into class labels. Since early stage of using ML for material discovery, classic algorithms such as linear regression, supported vector machine and Bayesian inference methods are adopted. Although they are suitable for cases where the training dataset is small, they usually have limited capacity or require a good knowledge and presumption on the mapping to be learned. 

In the fashion of big data and deep learning, neural networks join in as one type of widely used ML models. While they have high representative capacity so that any function can be approximated in principle, massive training samples are needed, or otherwise, the learned mapping has poor generalization ability in domains that are not abundantly sampled. Another critique is the black-box nature of neural networks, which hinders drawing physical insights. A step improving the interpretability is taken in Ref.~\cite{Xie2018}. by constructing specific graph convolutional neural networks for periodic crystals, so that the structural information is apparent. As the PPM is only a property value calculator, a ML material discovery/design scheme usually includes other components such as strategy of designating the search space, DFT calculation and even experimental verification~\cite{Ward2016,Raccuglia2016,Xue2016}.

In PAM we start from a microscopic model/description of certain properties and represent the atoms as vectors (usually in high dimensional Euclidean space) so that a relation between the representative vectors and the property can be constructed, as shown in Fig.~\ref{rationale}(b). The orientation and/or length of the vectors are optimized to make the model's output a good approximation of the property values. As long as the microscopic model is made of continuous functions, a property-specified similarity is naturally and quantitatively defined by the distance among the vectors.  Based on the similarity we can use the atom substitution as a recommendation strategy to designate the search space for DFT or experimental verification.

As shown at the output end in Fig.~\ref{rationale}(b), the aimed property of PAM is not necessarily the targeted property that we directly inspect, but relevant properties that are easier for modelization and data collection, say, formation energy vs. thermal stability and binding energy vs. mechanical strength. From property selection to modelization, PAE is built on physical insight and understanding. Compared to PPM, this is somewhat a drawback in the sense that the microscopic models need some endeavour to be set up. However, the endeavour pays off. Besides inherently being a material recommender, the built-in interpretability facilitates a utility much peculiarly pertinent to PAM, which is our focus in the remaining.

In Ref.~\citep{Zhou2017} and~\cite{Jaeger2018}, the authors embedded atoms and organic groups, respectively, as vectors according to their cooccurrence statistics. Their results confirm the validity of periodic table grouping and known functional similarity among the organic groups. While these works suggest that the embedding approach can learn general knowledge, their further usage of the embedded vectors as input for PPM dose not add up much to our arsenal. Since output of each hidden layer in a neural network is a sort of property embedding, even using the vanilla chemical formula as input, embedding of the constitutional information is implemented by the neural network. From this perspective, the embedding based on cooccurrence amounts to an input preprocessing, which can be beneficial for numerical stability and convergence but does not touch the black-box issue or mitigate the demand of sample abundance.

While still maintaining the ability of distilling general knowledge, when the atoms are embedded with the selected properties as aim, the learned knowledge is not too general to go beyond well known facts. Moreover, since the selected properties have been endowed with our physical understanding, we do not need to forcefully construct a black-box mapping between these properties and the targeted property. Then a simple algorithm that does not demand massive learning samples can be effective enough. The two traits render PAE an applicable tool in the situation where we have abundant data to learn general knowledges but the data provision is relatively scarce for the specific problem of interest to us. 

Study in high entropy alloys (HEAs)~\cite{Ye2016,Miracle2017,Lederer2018} is a proper arena to showcase this utility. Although we have a large number of binary and ternary alloy entries in widely used databases, their data for quartet and beyond are rare or null. Meanwhile, the periodic table grouping are less pertinent for metals, because, unlike valence or ionic bonds, electronic states in metals are highly delocalized so that the stability is much less dependent on atomic structure of neighbouring atoms. This is the reason why the embedding based on cooccurrence is ineffective for transition metals, which present in most of HEAs. More specifically, here we address the problem of searching for single solid-solution phase HEAs~\cite{Zhang2008,Troparevsky2015,Tan2018,Tian2017,Wang2015,Tapia2018,Gao2013,Guo2013,Yang2012,Ye2015,Chattopadhyay2018}, which is a meaningful problem in itself, and previous works of which can be used as comparison benchmarks.

\begin{figure}
\includegraphics[width=0.9\textwidth]{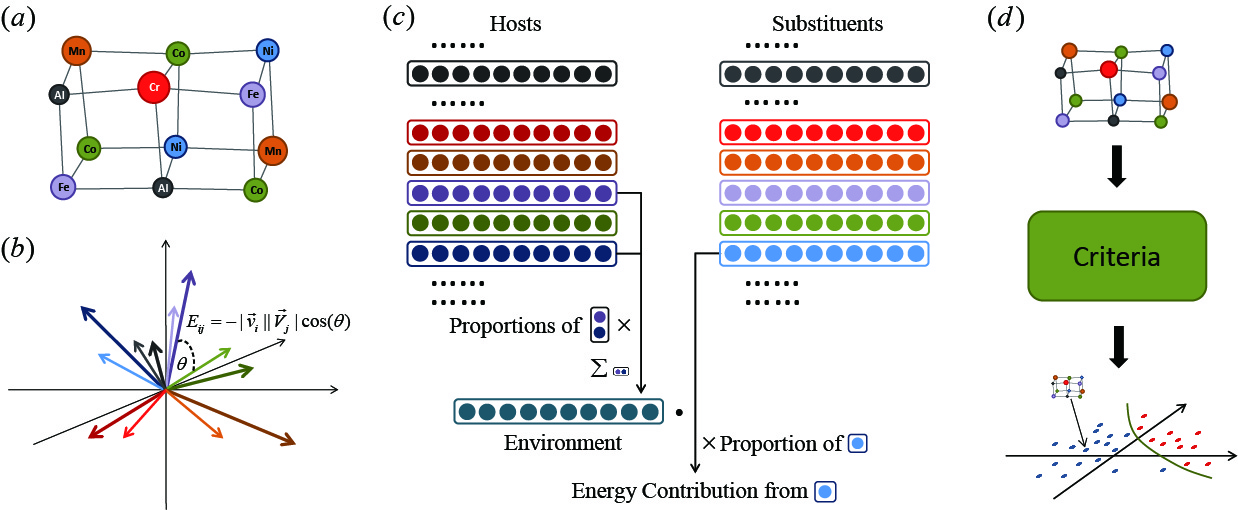}
\caption{(a) Motivation of property selection: lattice distortion could cause variation of energy. (b) Each atom is represented both as host and substituent vectors, and binding energy is defined between host and substituent as the negative inner product. (c) Energy contribution of an ingredient is summation of its interaction with the environmental hosts weighted by their proportions. (d) After the optimization, criteria separating the different phases are built based on learned knowledge.}
\label{frame}
\end{figure}

In Fig.~\ref{frame} are schematics of our motivation of property selection, vector representation scheme of atoms, the microscopic description and a final classification procedure. Let us go through them one at a time. In Ref.~\cite{Zhang2008} and~\cite{Troparevsky2015}, the authors showed respectively, that closeness in binary enthalpy of the ingredient atoms and similarity in atom size are nice properties to predict whether an HEA is a single phase alloy. Our choice of energy per atom as the embedding aim is motivated by the intuition that size mismatch should also be reflected in the energy since it kind of stretches the lattice~\cite{Wang2015} and could cause energy variation. In addition, energy is closely related to enthalpy in that enthalpy at $T=0$ K is equal to the formation energy. Regarding the two observations, energy per atom could be a property pertinent to our problem, and at the same time, is well recorded in most databases.

Unlike the embedding based on cooccurrence statistics where each entity corresponds to one vector only, we assign two vectors to each atom involved---the host vector and the substituent vector. The reason is more a technical predicament than the physical picture that we generate new materials by atom substitution. As the first step to relate the representative vectors to the aimed property---energy, we define \textit{binding energy} between two ingredients as the inverse of distance. Here, the inverse is used, because for stable matters, total energy $E<0$ (so is the energy per atom), and we wish the atoms with stronger binding tendency to lie closer. Then, if the Euclidean distance is used, positive energy is excluded at all. 

A more proper distance is the cosine distance, i.e. the inner product. The inequality $(\vec{A}-\vec{B})\cdot(\vec{A}-\vec{B})>0$, however, implies that binary matters would always have positive formation energy if atoms are only represented by a single vector, because now $(\vec{A}-\vec{B})\cdot(\vec{A}-\vec{B})$ is exactly the formation energy of binary matter AB. This implies that alloys can not stably exist, which clearly contradicts with the reality. Single vector representation is unviable. We can technically circumvent the predicament by the dual vector assignment, and define the binding energy between atoms A and B as inner product of host vector $\vec{A}$ and substituent vector $\vec{b}$, as sketched in Fig.~\ref{frame}(b). 

Now, for two atoms A and B, we have four inner products $\vec{A}\cdot\vec{a}$, $\vec{B}\cdot\vec{b}$, $\vec{A}\cdot\vec{b}$ and $\vec{B}\cdot\vec{a}$. They are related, but due to the separation of host and substituent, formation energy of binary alloys are not restrict to be positive. With the dual assignment, both $\vec{A}\cdot\vec{b}$ and $\vec{B}\cdot\vec{a}$ have the meaning of binding energy between atoms A and B. Here, we do not impose $\vec{A}\cdot\vec{b}=\vec{B}\cdot\vec{a}$. In other words, instead of the real binding energy, we implement evaluation for a less physical property: how much a substituent atom contributes to the totally energy with certain atoms presented in its neighbourhood as host.

We exemplify our formulation of energy per atom illustrated in Fig.~\ref{frame}(c) through a ternary matter with chemical formula A$_{x}$B$_{y}$C$_{z}$, which reads
\begin{numcases}{}
E_{per} = \frac{1}{2} \sum_{i=ABC} w_i \vec{v}_i\cdot \vec{N}_{i},\label{linear}\\
w_A = \frac{x}{x+y+z},\\
\vec{N}_{A} = \frac{(x-1)\vec{A}+y\vec{B}+z\vec{C}}{x-1+y+z},\label{eviron}
\end{numcases}
where $\vec{v}_i$ denotes the substituent vector, $w_A$ is the proportion of $A$, and $\vec{N}_{i}$ the environment vector representing the host atoms around. Factor $1/2$ is included since $\vec{v}_i$ is only one end of the substituent-host interaction. Similar to ternary, the case of other numbers of ingredients can be readily inferred. One point we should note is that calculating the ratio of hosts in the environment vector as in Eq.~(\ref{eviron}) implies exclusion of the atom itself, which is different with the usually used regular solution assumption in HEA studies. We such formulate it since the data in the material project database~\cite{Jain2013} are from DFT calculations in which the crystal structure and atom positions are fixed. When the optimized results are applied to HEAs, we assume they are regular solutions and do not perform the self-exclusion.

To avoid unconscious data selection, we will use the experimental data collected in Ref.~\cite{Ye2016} for our problem of HEA phase separation. To cover the elements presented there, we include $45$ metals and nonmetals P and Si in our embedding (refer to supp. for the full list). There are $10075$ materials in the database that are composed by these elements. Neglecting structural influence, we use the one with lowest energy among materials under the same chemical formula. After this filtering, $7396$ matters remains: $2490$ binaries, $4835$ ternaries and $24$ quartets. During our early calculation, we found the numerical stability is poor in the sense that the results clearly vary due to randomness in initialization and stochastic gradient descent. It appears that Eq.~(\ref{linear}) can not represent complex interactions due to its linearity. For the reason, we leave the ternaries as correction to be addressed later, and firstly optimize the embedding using the singlet and binary data only. We use the mean squared error as the loss function,
\begin{equation}
L = \sum_{i} \frac{(E_{per}^i - E_{d}^i)^2}{S},
\end{equation}
with summation over the training data. Here, $E_{d}^i$ is the recorded energy per atom in the database, and $S$ is the number of training samples.

\begin{figure}
\includegraphics[width=0.9\textwidth]{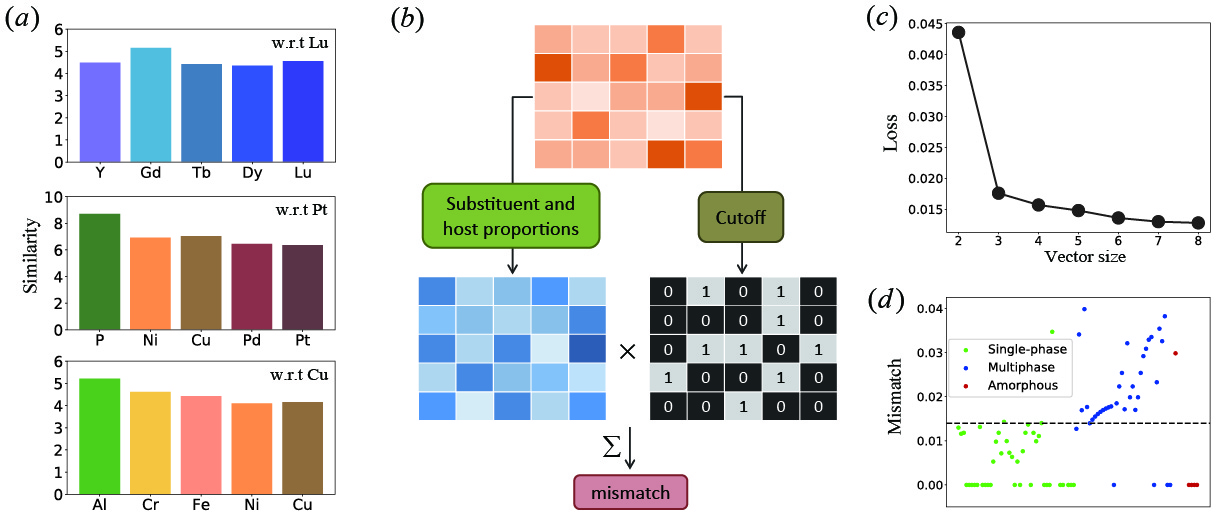}
\caption{(a) Single-phase alloys usually have higher similarity among its components than the other two sorts, and one element with strong binding tendency to the others is the major contributor of structure distortion. (b) An energetic mismatch for a material is evaluated as a weighted summation over similarities that exceed a cutoff. (c) The vector size is determined to be Dim=$3$, which corresponds to the critical point in the loss. (d) With the cutoff $0.7459$ and the borderline at $0.0144$, the ration of correctly separated samples is $76/90$.}
\label{criteria}
\end{figure}

The size (dimension) of the vectors is a hyper-parameter, which is usually determined with validation dataset. Because we do not have one, we use another strategy to prevent overfitting---critical point in the loss. We will present results with Dim $=3$, since from Fig.~\ref{criteria}(c) we see that over it the loss decreases slowly. Another justification is geometrical intuitive. When Dim $\leq2$, the vectors (actually scalars for Dim $=1$) are enforced an order in the position on a line for Dim $=1$ and in the polar angle for Dim $=2$, which implies that a vector can only have two neighbours. For Dim $\geq3$ the number of neighbours can be infinite, which is an qualitative transition for freedom of arranging the vectors. Another technique detail is the initialization values. To facilitate numerical stability, we advice to start with small values, so that the vectors can effectively adapt their orientation at the early stage of training. If the initial vectors are long, it is more likely to stuck in suboptimal traps. In this work, normal distribution with standard error $0.01$ is used for the vector elements.

After optimization, we can define an energetic similarity with respect to each host atom. Ternary A$_{x}$B$_{y}$C$_{z}$, for instance, has the vector $(\vec{A}\cdot\vec{a}, \vec{A}\cdot\vec{b}, \vec{A}\cdot\vec{c})$ as similarity among A,B,C w.r.t host A and the other two w.r.t B and C. In Fig.~\ref{criteria}(a), we present representative results for good single-phase HEAs in the top panel where the ingredients are quite similar, and for typical amorphous HEAs in the middle where the atoms differ more. In the bottom panel is the similarities for the AlGrFeNiGu system, whose structural phase change from single-phase to multiphase with increase containment of Al. Our results well reflect this in that the Al is exactly the most dissimilar one. For a material, the full similarity metrics is a matrix (as sketched in Fig.~\ref{criteria}(b)) with rows and columns indexed by hosts and substituents, respectively. What are presented in Fig.~\ref{criteria}(a) are the last rows, and the other rows usually show similar pattern, except that the absolute values may vary a lot.

From Fig.~\ref{criteria}(a), we can have a rough impression that one ingredient with the strongest binding tendency (the longest bar) is usually the key player that distorts the lattice and cause structural inhomogeneity. In Fig.~\ref{criteria}(b), we show the procedure of promoting this observation to a quantitative criterion. First, the issue of absolute value variation among different rows is addressed by row-wise normalization in which the row maximum is normalized to unity. This results in the orange matrix on the top of Fig.~\ref{criteria}(b). To proceed, a relation between similarity and structural phase is needed, for which we simply assume a cutoff, the only parameter in our criterion. When the similarity is above the cutoff, we consider the energetic mismatch with the key player is ineffective and do not count it in, otherwise, the player scores $1$. By this we have the white-black matrix on the right. To reflect effect of atom proportions, elements of the normalized matrix are weighted by the corresponding host and substituent proportions to arrive the blue matrix on the left. Finally, a mismatch value is obtained for each material by summarizing the element-wise product of the two matrices.

The optimal cutoff value is the one that results in best separation of the single-phase from the other two phases. As shown in Fig.~\ref{criteria}(d), with cutoff $0.7495$ and a borderline of mismatch at $0.0144$, we can have a seperation score $76/90$. This means that the number of single-phase HEA assigned to multiphase or amophous HEA, and the number of misassignments the other way around are $14$ in total. We note that due to the cutoff nature, the optimal cutoff is actually a narrow range around $0.7495$, since within it the results do not change. In Ref.~\cite{Tan2018} the authors compared major empirical criteria using the same materials. Compared to them, our separation is not the best, but no worse than most of those criteria. Considering the simplicity of our criterion, this comparability is an indication of the viability of our approach.

Now let us deal with learning knowledge from the ternaries (we neglect the 24 quartets for computational convenience). Because the weights are not variables and Eq.~(\ref{linear}) is linear, the number of inner product is the number of independent degree of freedoms we can have. $M$ substituent-host pairs can only offer $M^2$ freedoms, regardless of the vector size. When the complexity requires more freedom to represent it, increasing vector size is ineffective. We could increase the vector size and at the same time add nonlinear operations before taking the inner products. Based on our attempts, however, while this can fit the output well to the recorded energy, it is a mess when the we use the optimized vectors for HEA phase separation. In that way, we actually take the step that lead to neural networks' black-box nature and weak generalization ability.

\begin{figure}
\includegraphics[width=0.43\textwidth]{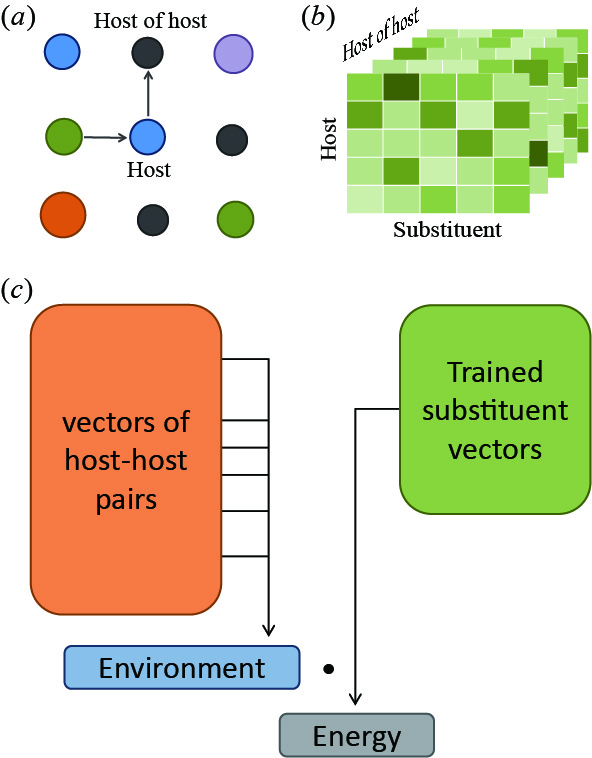}
\caption{(a) Weights of the host-host pairs are defined as product of proportions of the two hosts, for which the indirect host is treated as host of the host. The arrows mean \textit{is hosted by}. (b) The similarity metrics now is a 3D array for each matter. (c) The perturbation like procedure is generally similar to that in Fig.~\ref{frame}(c), except an important distinction that the embedding vectors of substituents have been optimized previously.}
\label{perturb}
\end{figure}

Another strategy to cope with complexity is increasing the number of entities to be embedded, which increases the number of freedom. As an attempt, we treat the host-host pairs as objects to be embedded, whose proportion is defined as product of their individual proportions. As shown in Fig.~\ref{perturb}(a), when calculate the proportion of a host-host pairs, the indirect host is regard as the host of the direct host, which means we do not exclude the substituent indirectly hosted by it. Take A$_{x}$B$_{y}$C$_{z}$ for example, the weight of  pair B$\rightarrow$A in the environment of $A$ is
\begin{equation}
w_{B\rightarrow A} = \frac{y}{x-1+y+z}\times\frac{x}{x+y-1+z},
\end{equation}
where the arrow means \textit{is hosted by}, and $B\rightarrow A$ and $A\rightarrow B$ is treated as two distinct entities. To facilitate the numerical stability, we do not redo the embedding from scratch, but consider the knowledge from ternaries as something like perturbation. This is quite beneficial, because as shown in Fig.~\ref{perturb}(c), we now keep the substituent vectors unchanged, which are anchors for the other half of the inner product.

Besides this perturbation like aspect which we would like to emphasis, there are two other technical differences with the original calculation. Being treated as perturbation, our targeted value is not the recorded energy per atom, but $\Delta E = E_d-E_{per}$, with $E_{per}$ the per atom energy calculated based on the previously optimized vectors. In addition, the similarity metrics is expanded from a matrix to a 3D array, where the depth dimension represents the hosts of hosts. Its element can be understand as binding energy between a substituent and a host modified by the host of host. Accordingly, its value is $E_{ijk}=E_{ij}+\Delta E_{ijk}/2$, where $E_{ij}$ is the unperturbed binding energy between atom $i$ and host $j$, as those in Fig.~\ref{criteria}(a), and $\Delta E_{ijk}$ is the binding energy between substituent $i$ and host-host pair $j\rightarrow k$.  Then following the procedure in Fig.~\ref{criteria}(b), the cutoff is determined as $0.7353$ and the borderline is at $1.678E^{-3}$. The separation result is similar to that in Fig.~\ref{criteria}(d), with an improvement from $76/90$ to $79/90$.

Compared to the empirical criteria, ours does not work well in separating out amorphous alloys. Many factors can contribute to the inaccuracy. Most importantly, the simple criterion of energetic mismatch can not fully capture the complexity. One can note form Ref. that the amorphous alloys usually have very low enthalpy. The atoms' strong tendency to bind up may require stronger energetic similarity to keep the structural homogeneity. In other words, the cutoff value should vary matter-by-matter according to their enthalpy. It could also be because the cutoff criterion instead of a continuous function is a very rude division. But the other way around, it is this simplicity that evinces the validity of our approach, since it avoids the possibility that complicated criteria can easily become a game of fitting. Neglecting structural information is another issue. Although the chemical formula contains some structural information, accurate number of nearest and next nearest neighbour atoms varies with crystal structure. Our evaluation for the weights is somewhat rough. 



Before conclusion, we remark on the essence of the embedding method in this work. Since our microscopic model is an expression that represents total energy as summation of the parts, one can consider our embedding approach as a way of extracting the binding energies. However, the collectively defined binding energy is different with the original DFT data of the equal atomic binary alloys in several aspects. First, as discussed above, due to the dual assignment their meaning are inequivalent by definition. The effectiveness of our results is a reminder that what really needed is not necessarily a well defined physical model that links fundamental laws to experimental observation, but basically, a data analysis model motivated by microscopic understanding. Second, since many binary relations (say, between Cu and Fe, Co, Ni) are null in the database due to instability of the alloys composed by the two elements, these values are sheerly inferred through the binding of the two elements with the rests. Actually, the other energies also posses such a sense of being inferred, since each embedded vector is collectively determined by its direct of indirect interaction with all the others. This is favourable for our problem in that HEAs consist of multiple elements with varying quota.

In summary, we showcased usage and utility of PAE through separating the single-phase alloys with those in multiphase and amorphous phase. Implication and outlook of our work lie at three levels. For the HEA phase separation problem, it is an example of one property only. We can include other quantities such as variation in density and crystal constants to reflect the lattice stretch, so that the problem can be addressed with multiple criteria. More specifically to our model, one way of representing the structural information is treating the same atom in different structure as separate vectors. While the simplicity of the criterion is fine for us to show viability of PAM, it is not necessary for practical tasks. As mentioned above, some thermodynamic information may be incorporated in.

As for other material discovery tasks, in principle once a microscopic model is given, a PAE scheme could be constructed for the involved entities, so it is generable and flexible. From classical to quantum mechanics, a system can be represented by its Hamiltonian, which implies that energetic representation is an important aspect for physical microscopic models. Our dual assignment scheme can be a general strategy for these application to avoid the problem of positive defined formation energy. Due to its ability of maintaining numerical stability and adding up on a basis, we believe that the perturbation like procedure can be pertinent to and useful for other problems, in the regard that perturbation techniques are applied by physicists to a wide range of problems.

This sort of ***2vec models was originally proposed~\cite{Mikolov2013a,Mikolov2013b} and is widely used in natural language processing under the name \textit{word2vec}. Since a language is a system that is built upon practice and convention, its evolving and form are not governed by fundamental laws as those in physical sciences. For this reason, previous usage of the models has been focused on reflecting similarity in statistical patterns. At the inter-discipline level, our work shows the possibility and viability of the application in situations where microscopic mechanism is in play. As reductionism is a paradigm of understanding physical and sociological phenomena, and vectorization is a general scheme of quantifying entities and concepts, the idea of PAE can be applicable far beyond usage in material discovery.

\bibliography{references}

\end{document}